\documentstyle[pra,aps,twocolumn,floats,psfig]{revtex}

\begin{document}

\draft

\title{Atomic dynamics in evaporative cooling of\\ 
trapped alkali atoms in strong magnetic fields}

\author{O. H. Pakarinen and K.-A. Suominen}

\address{Helsinki Institute of Physics, PL 9, FIN-00014 Helsingin yliopisto,
Finland}

\date{\today}

\maketitle

\begin{abstract}
We investigate how the nonlinearity of the Zeeman shift for strong magnetic 
fields affects the dynamics of rf field induced evaporative cooling in magnetic
traps.  We demonstrate for the $^{87}$Rb and $^{23}$Na $F=2$ trapping states 
with wave packet simulations how the cooling stops when the rf field frequency 
goes below a certain limit (for the $^{85}$Rb $F=2$ trapping state the problem 
does not appear). We examine the applicability of semiclassical 
models for the strong field case as an extension of our previous work 
[Phys. Rev. A {\bf 58}, 3983 (1998)]. Our results verify many of the aspects 
observed in a recent $^{87}$Rb experiment [Phys. Rev. A {\bf 60}, R1759 (1999)].
\end{abstract}


\narrowtext

\section{Introduction}

\begin{figure}[htb]
\noindent\centerline{
\psfig{width=90mm,file=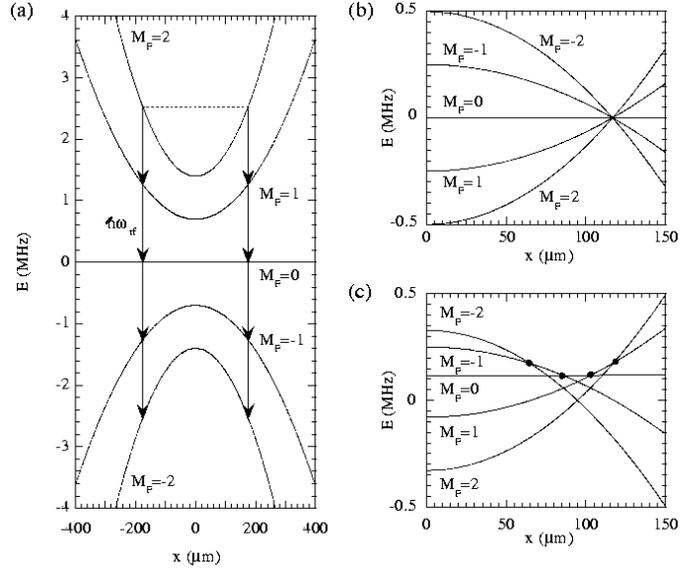}
}
\vspace*{0mm}
\caption[f1]{The magnetic trap potentials for spin-polarised $^{87}$Rb ($F=2$).
(a) The spin flips that lead to evaporation are achieved by an rf field induced
multistate transition at a specific spatial distance from the trap center.
Here $B_0=0.0001$ T. (b) In the curve crossing description the resonances
appear as degeneracies. Here $B_0=0.0001$ T. (c) For strong fields the 
multistate crossing transforms into a sequence on two-state crossings between 
adjacent $M_F$ states. Here $B_0=0.0020$ T. The circles mark those
crossings where the involved adjacent $M_F$ states are also coupled.
In (b) and (c) we have $\nu_{\rm rf}=0.25$ MHz+$\nu_0$.
\label{fig1}}
\end{figure}

Bose-Einstein condensation of alkali atoms in magnetic traps was first observed
in 1995~\cite{Anderson95}, and since then the development in related research
has been been very swift. Typically the hyperfine state used in the alkali
experiments is the $F=1$ state, although condensation has been demonstrated for
the $^{87}$Rb $F=2$ case as well~\cite{Jila}. The trapping of atoms is based on
moderate, spatially inhomogeneous magnetic fields, which create a parabolic,
spin-state dependent potential for spin-polarised atoms, as shown in
Fig.~\ref{fig1}(a). For slowly moving atoms the trapping potential depends on
the strength of the magnetic field $B$ but not on its
direction~\cite{Suominen98}. In practice the field is dominated by a
constant bias field $\vec{B}_{\rm bias}$, which eliminates the Majorana spin 
flips at the center of the trap. 

In evaporative cooling the hottest atoms are removed from the trap and the
remaining ones thermalise by inelastic collisions. This leads to a decrease in
temperature of the atoms remaining in the
trap~\cite{Hess86,Tommila86,Ketterle96}.  Continuous evaporative cooling
requires adjustable separation into cold and hot atoms. This is achieved by
inducing spin flips with an oscillating (radiofrequency) magnetic field, which
rotates preferably in the plane perpendicular to the bias
field~\cite{Ketterle96,Pritchard88}. In the limit of linear (weak) Zeeman
effect the rf field couples the adjacent magnetic states $M_F$ resonantly at
the spatial location determined by the field frequency [Fig.~\ref{fig1}(a)].
Hot atoms oscillating in the trap can reach the resonance point and exit the
trap after a spin flip to a nontrapping state. Using the rotating wave
approximation we can eliminate the rf field oscillations, and obtain the curve
crossing description of resonances [Fig.~\ref{fig1}(b)].

\begin{figure}[htb]
\noindent\centerline{
\psfig{width=90mm,file=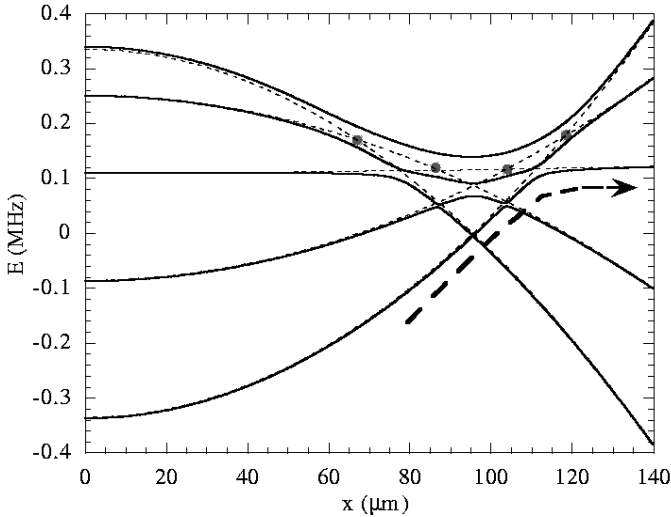}
}
\vspace*{0mm}
\caption[f2]{The adiabatic potentials (solid lines) and bare $M_F$
states (dotted lines) for $^{23}$Na at $B_0=0.0010$ T, with
rf field coupling $\Omega= (2\pi) 20$ kHz. The arrow indicates the
semiadiabatic process for transfer from the $M_F=2$ state to the
$M_F=0$ state. Alternatively one can describe the process as an
off-resonant two-photon transition.
\label{fig2}}
\end{figure}

The dynamics of atoms as they move past the resonance point can be described
with a simple semiclassical model~\cite{Vitanov97}, which has been shown to
agree very well with fully quantum wave packet calculations~\cite{Suominen98}.
The model, however, can be applied only if the resonances between adjacent
$M_F$ states occur at the exactly same distance from the trap center. When
the nonlinear terms dominate the Zeeman shifts, the situation changes, as shown
in Fig.~\ref{fig1}(c). The adjacent resonances become separated and one
expects to treat the evaporation as a sequence of independent Landau-Zener
crossings as suggested by Desruelle et al. in connection with their recent
$^{87}$Rb experiment~\cite{Desruelle99}. We show that there is an intermediate
region where off-resonant two-photon transitions from the $M_F=2$ state to the
$M_F=0$ state, demonstrated in Fig.~\ref{fig2}, play a relevant role. 

In general there is a competition between the adiabatic following of the
eigenstates (solid lines in Fig.~\ref{fig2}), which leads to evaporation, and
nonadiabatic transitions which force the atoms to stay in the trapping
states. In $^{23}$Na the nonadiabatic transitions can lead to highly
inelastic collisions~\cite{Suominen98}. 

In the experiment by Desruelle et al. it was found that for a strong bias field
the nonlinear Zeeman shifts remove some resonances completely, thus making it
impossible to make a spin flip to a nontrapping state. Our calculations
confirm this observation. We also show that although evaporation could
continue via off-resonant multiphoton processes, such a process is not
practical. The stopping of evaporation at some finite temperature occurs for
the $^{87}$Rb and $^{23}$Na $F=2$ trapping states, but not e.g.~for the
$^{85}$Rb $F=2$ trapping state.

In Sec.~\ref{basis} we write down the formalism for the Zeeman shifts and
show the basic properties of the field-dependent trapping potentials.
We describe the fully quantum wave packet approach and corresponding
semiclassical theories in Sec.~\ref{approach}, present and discuss the results
in Sec.~\ref{results}, and summarize our work in Sec.~\ref{conclusions}. 

\section{The Zeeman structure}\label{basis}

\subsection{$^{23}$Na and $^{87}$Rb}

The Zeeman shifts can not be derived properly in the basis of the hyperfine
states (labelled by $F$ and $M_F$)~\cite{Bransden83,Rose95,Pethick97}. 
We need to consider the atom-field Hamiltonian in the $(I,J)$ basis: 
\begin{equation}
   H = A\vec{I}\cdot\vec{J}+CJ_z+DI_z,  \label{H}
\end{equation}
where $\vec{I}$ and $\vec{J}$ are the operators for the nuclear and total
electronic angular momentum, respectively. The first term describes the
hyperfine coupling; $E_{\rm hf}=h \nu_{\rm hf}=2A$, where $E_{\rm hf}$ is the
hyperfine splitting between the $F=1$ and $F=2$ states. Here $\nu_{\rm hf}=
1772$ MHz for $^{23}$Na and $\nu_{\rm hf}=6835$ MHz for $^{87}$Rb. 

The magnetic field dependence arises from the two other terms, with
$C=g_J\mu_BB$ and $D=-\alpha\mu_NB$, where the Bohr magneton is $\mu_B =
e\hbar/2m_e$, the nuclear magneton is $\mu_N=e\hbar/2m_p$, and the Lande factor
is $g_J=2$. Here $\alpha= 2.218$ for $^{23}$Na and $\alpha=2.751$ for $^{87}$Rb.
But $\mu_B/\mu_N\sim 1000$, and in fact we can omit the third term in
Eq.~(\ref{H}).  

\begin{figure}[htb]
\noindent\centerline{
\psfig{width=90mm,file=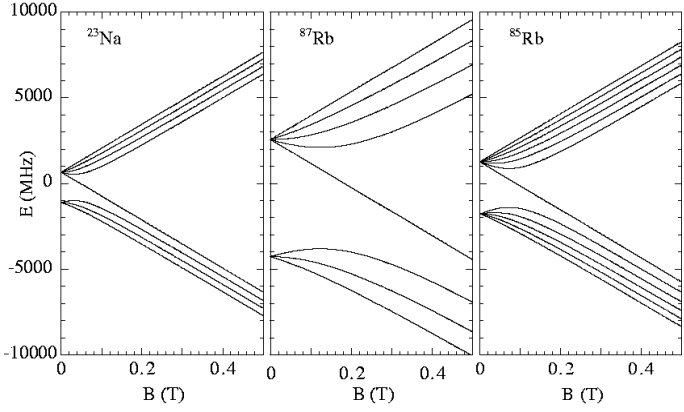}
}
\vspace*{0mm}
\caption[f3]{The Zeeman shifts for the ground state hyperfine
states of (a) $^{23}$Na, (b) $^{87}$Rb and (c) $^{85}$Rb. Note
that the situations considered in this paper take place in a region located
very close to $B=0$ in the scale of these pictures. 
\label{fig3}}
\end{figure}

For $^{23}$Na and $^{87}$Rb we have $I=3/2$ and $J=1/2$ (leading to $F=1$ or
$F=2$ with $\vec{F}=\vec{I}+\vec{J}$). Our state basis is formed by the
angular momentum states labelled with the magnetic quantum number pairs
$(M_I,M_J)$. When we evaluate the matrix elements of $H$ [using the relation
$\vec{I}\cdot
\vec{J}=I_zJ_z+\frac{1}{2}(I_+J_-+I_-J_+)$], the states that correspond to the
same value of $M_F=M_I+M_J$ form subsets of mutually coupled states. By
diagonalising the Hamiltonian we obtain its eigenstates. The states which
correspond to the $F=2$ state in the $B\rightarrow 0$ limit (labelled with
$M_F$) have the energies $E_{\rm M_F}$:
\begin{eqnarray}
   E_{+2}&=&\frac{1}{2}C,\nonumber\\
   E_{+1}&=&\frac{1}{2}\sqrt{4A^2+2AC+C^2}-A,\nonumber\\
   E_{0} &=&\frac{1}{2}\sqrt{4A^2+C^2}-A,\label{E}\\
   E_{-1}&=&\frac{1}{2}\sqrt{4A^2-2AC+C^2}-A,\nonumber\\
   E_{-2}&=&-\frac{1}{2}C.\nonumber
\end{eqnarray}
These energies have been normalised to the energy of the $F=2$ state for
$B=0$. In Fig.~\ref{fig3}(a) and (b) we show the Zeeman shifts for all 
hyperfine ground states of $^{23}$Na and $^{87}$Rb, but normalised to the
ground state energy in the absence of hyperfine structure. For small magnetic
fields ($B\ll$ 1~T) we get 
\begin{equation}
   E_{\rm M_F}\simeq E_{\rm hf}[\varepsilon M_F+(4-M_F^2)\varepsilon^2],
\end{equation}
where $\varepsilon= \mu_BB/(2E_{\rm hf})$. In terms of $F$ and $M_F$ the
linear Zeeman shift is $E_{\rm M_F}=g_F\mu_BBM_F=E_{\rm hf}\varepsilon M_F$ as
the hyperfine Lande factor is $g_F=1/2$.

\begin{figure}[hbt]
\noindent\centerline{
\psfig{width=90mm,file=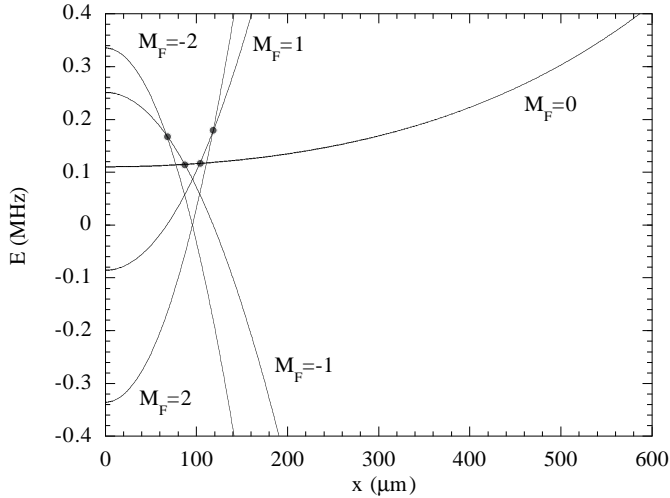}
}
\vspace*{0mm}
\caption[f4]{The $M_F=0$ state becomes a trapping state very
quickly as $B$ increases. Here we show as an example the situation for
$^{23}$Na when $B_0=0.0010$ T. The oscillation frequency for this state is
naturally much smaller than for the other trapping states.
\label{fig4}}
\end{figure}

The necessary condition for evaporation is that the rf field induces a
resonance between the states $M_F=2$ and $M_F=1$. The location of this
resonance defines the division between the hot and cold atoms. By decreasing
the rf field frequency $\nu_{\rm rf}$ we both move the resonance point closer 
to the trap center as well as allow more atoms to escape the trap. For
small $B$ fields all adjacent states are resonant at the same location for
any $\nu_{\rm rf}$. But in case of strong magnetic fields, typically larger 
than about 0.0002 T, due to the nonlinear Zeeman shifts the resonances
separate. Furthermore, the other resonances than the $M_F=2-M_F=1$ one 
in fact move towards the trap center faster, and reach it while the
$M_F=2-M_F=1$ resonance still corresponds to some finite temperature. 
When $\nu_{\rm rf}$ is lowered further, the other resonances begin to 
disappear.  At strong $B$ fields the $M_F=0$ state is also a trapping state, 
as shown in Fig.~\ref{fig4}, so for effective evaporation one really needs to 
reach the $M_F=-1$ state.

\begin{figure}[htb]
\noindent\centerline{
\psfig{width=90mm,file=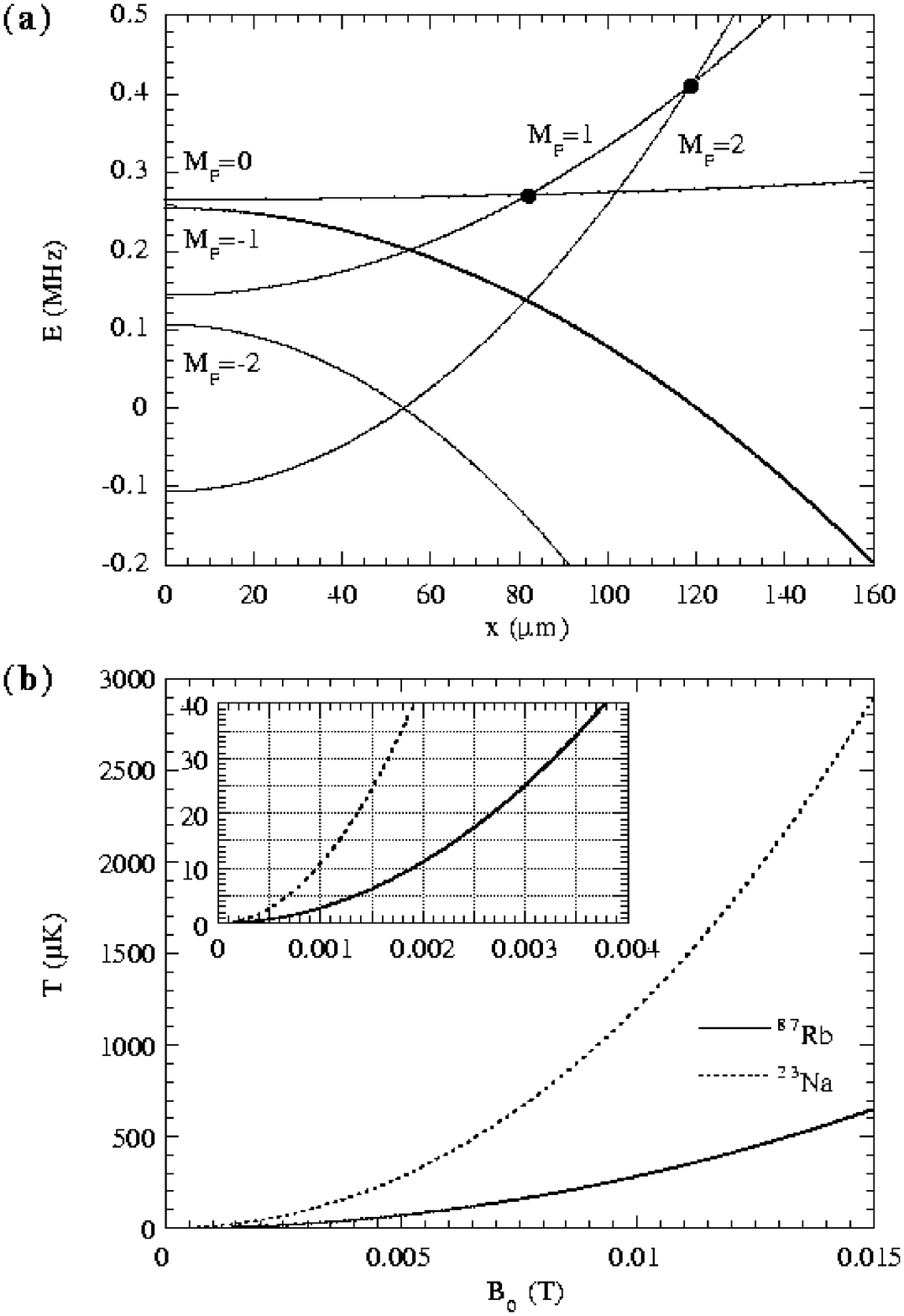}
}
\vspace*{0mm}
\caption[f5]{(a) For $B=0.00155$ T and $\nu_{\rm rf}=$0.25 MHz+$\nu_0$ the
resonance between the $M_F=0$ and $M_F=-1$ states for $^{23}$Na can not be
achieved ($B_{\rm cr}=0.00152$~T). The possibility for a tunnelling-like 
transfer (which could also be called off-resonant process via power broadening)
exists, though. (b) The lowest energy (in temperature units)
for which evaporation is allowed as a function of trap center field $B_0$
for $^{87}$Rb and $^{23}$Na with the trap parameters used in this paper.  
\label{fig5}}
\end{figure}

At the critical frequency $\nu_{\rm cr}$ the crossing between the states 
$M_F=-1$ and $M_F=0$ disappears. Alternatively, for a fixed frequency 
$\nu_{\rm rf}$ we have a critical value $B_{\rm cr}$ for the $B$ field; 
the resonances disappear when $B\gtrsim B_{\rm cr}$ (for practical reasons 
we have chosen to modify $B$ rather than $\nu_{\rm rf}$ in our wave packet 
studies). In Fig.~\ref{fig5}(a) we show the potential configuration when 
$\nu_{\rm rf}$ is slightly below $\nu_{\rm cr}$. Since $\nu_{\rm cr}$ 
corresponds to the state separation at the center of the trap, it is 
independent of the trap parameters such as the trap frequency. 

For a specific trap configuration $\nu_{\rm cr}$ can be 
converted into a minimum kinetic energy required for
reaching the resonance between the states $M_F=2$ and $M_F=1$. In
Fig.~\ref{fig5}(b) we show this minimum kinetic energy in units of temperature
as a function of magnetic field strength for $^{23}$Na and $^{87}$Rb, and for
the trap configuration used both in our simulations and in the experiment by
Desruelle et al.~\cite{Desruelle99}.

In the intermediate region $0\ll B\lesssim B_{\rm cr}$, 
where the necessary crossings exist but are 
separated, the processes take place via two possible routes. We can have
off-resonant multiphoton processes, that e.g.~lead to adiabatic transfer from
the $M_F=2$ state to the $M_F=0$ state. This example is demonstrated in
Fig.~\ref{fig2} where we show also the eigenstates of the system, i.e., the
field-dressed potentials. When the relevant resonances are well separated, the
evaporation takes place via a complicated sequence of crossings, as indicated
in Fig.~\ref{fig1}(c). This will be demonstrated with wave packet simulations
in Sec.~\ref{results}.

\subsection{$^{85}$Rb}

For the isotope $^{85}$Rb we have $I=5/2$ and $J=1/2$, so the ground state
hyperfine states are $F=2$ and $F=3$, as shown in Fig.~\ref{fig3}(c). Now
the $F=2$ trapping state is the lower hyperfine ground state. Thus the behavior
of the $M_F$ states is different from the $^{87}$Rb and $^{23}$Na case. The
$B$ field dependence of the states related to $F=2$ is now
\begin{eqnarray}
   E_{+2}&=&\frac{3A}{2}-\frac{1}{2}\sqrt{9A^2+4AC+C^2},\nonumber\\
   E_{+1}&=&\frac{3A}{2}-\frac{1}{2}\sqrt{9A^2+2AC+C^2},\nonumber\\
   E_{0} &=&\frac{3A}{2}-\frac{1}{2}\sqrt{9A^2+C^2},\label{E85}\\
   E_{-1}&=&\frac{3A}{2}-\frac{1}{2}\sqrt{9A^2-2AC+C^2},\nonumber\\
   E_{-2}&=&\frac{3A}{2}-\frac{1}{2}\sqrt{9A^2-4AC+C^2},\nonumber
\end{eqnarray}
where now $E_{\rm hf}=3A$. For $^{85}$Rb we have $\nu_{\rm
hf}$ = 3036 MHz. Here the trapping states are now $F=2,M=-2$ and $F=2,M=-1$.
If we now define $\tilde{\varepsilon}=(2/3)\varepsilon=\mu_BB/(3E_{\rm
hf})$ we get approximatively
\begin{equation}
   E_{\rm M_F}\simeq -E_{\rm hf}[\tilde{\varepsilon}M_F
   +(9-M_F^2)\tilde{\varepsilon}^2].
\end{equation}
As $g_F=-1/3$, this agrees with the linear expression
$E_{\rm M_F}=g_F\mu_BBM_F=-E_{\rm hf}\tilde{\varepsilon} M_F$.

\begin{figure}[hbt]
\noindent\centerline{
\psfig{width=90mm,file=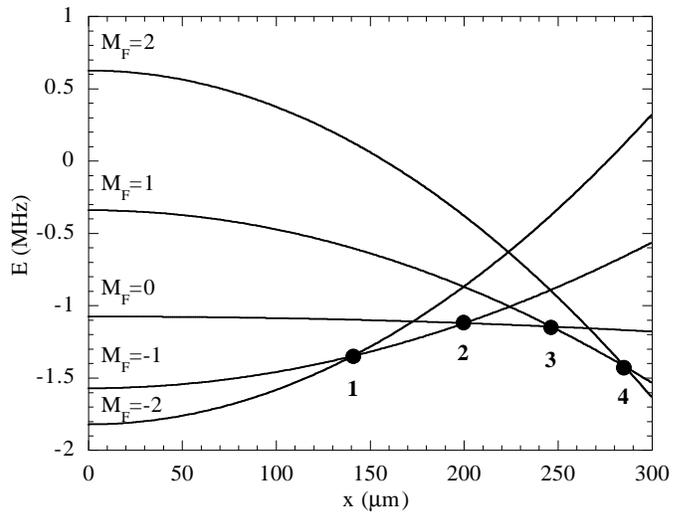}
}
\vspace*{0mm}
\caption[f6]{The effect of nonlinear Zeeman shifts on the evaporation from the
$^{85}$Rb $F=2$ state. The circles indicate for which crossing points the
couplings are nonzero. The numbers indicate the order in which an atom moving
out of the trap traverses the crossings. The kinetic energy required to leave
the trap is now determined by the difference between the trap center and the
second crossing. Here $B_0=0.0050$ T and $\nu_{\rm rf}=$0.25 MHz+$\nu_0$, with
$\nu_0=(E_{-2}-E_{-1})/h$.
\label{fig6}}
\end{figure}

The change of order in the $M_F$ state energy ladder means that with 
increasing $B$ field one never loses the crossing
points between the adjacent states. In other words, if we use an rf field that 
can couple the states $M_F=-2$ and $M_F=-1$ resonantly at some location $x_C$, 
then we always couple the rest of the states resonantly as well at distances 
larger than $x_C$. In Fig.~\ref{fig6} we see how this leads to a
sequence of crossings that allows hot atoms to leave the trap without the need
for sloshing. One must, however, take into account that the kinetic energy
required to leave the trap is now set by the difference between the energy of
the $M_F=-2$ state at the center of the trap, and the energy of the $M=0$ (or
$M_F=-1$) state at the point where the states $M_F=0$ and $M_F=-1$ are in 
resonance. In
other words, atoms need a kinetic energy equal or larger to the energy
difference between the trap center and the second crossing in Fig.~\ref{fig6}.

In this paper we limit our discussion on the $F=2$ case only, but it is
obvious that for the $^{85}$Rb $F=3$ trapping states we face the same problem
as in the $F=2$ case for $^{87}$Rb and $^{23}$Na. In general for the alkali
atoms we can expect that the problem will arise whenever we use the upper
ground state hyperfine state as the trapping state at strong $B$ fields.

\subsection{Trap configuration}

For simplicity we have assumed in our studies the same spatially inhomogeneous
magnetic field as in the experiment by Desruelle et al.~\cite{Desruelle99},
except that we have added a spatially homogeneous compensation field. 
This allows us to change the general field magnitude (depends on the bias 
field) while keeping the trap shape almost unchanged (depends also on the 
bias field). Thus we set~\cite{Desruelle99}
\begin{equation}
   B=B_0+\left(\frac{B'^2}{2B_{\rm bias}}
   -\frac{B''}{2}\right)(x^2+y^2)+B''z^2,\label{B}
\end{equation}
where $B'=9$ T/m, $B''/B_{\rm bias}=10^4$ m$^{-2}$, and the trap center field is
defined as $B_0\equiv B_{\rm bias}-B_{\rm comp}$. The actual trap is
cigar-shaped, which is a typical feature in many experiments. We have selected
the $x$ direction as the basis for our wave packet studies.  We set $B_{\rm
bias}= 0.0150$ T and use $B_{\rm comp}$ as a parameter to change $B_0$. 
Using $C=g_J\mu_BB$ with Eqs.~(\ref{E}) and~(\ref{B}) we get the spatially
dependent trapping potentials.

\section{Quantum and semiclassical models}\label{approach}

\subsection{Wave packet simulations}

For our wave packet studies we fix the rf field frequency to the value
$\nu_{\rm rf}=\nu_0+0.25$ MHz, where $\nu_0=[E_{+2}(x=0)-E_{+1}(x=0)]/h$.
With this setting the atoms need typically a kinetic energy about $E_{\rm
kin}/k_B\simeq24\ \mu$K in order to reach the crossing between the states
$M_F=2$ and $M_F=1$.  With our special definition of $\nu_{\rm rf}$ the
differences between $^{23}$Na and $^{87}$Rb appear mainly in the time scale of
atomic motion (Na atoms are lighter and thus move faster), and in scaling of
$B$. For our selected $\nu_{\rm rf}$ we have $B_{\rm cr}=0.00297$~T for
$^{87}$Rb and $B_{\rm cr}=0.00152$~T for $^{23}$Na. We have used the rf field 
strength $\Omega=(2\pi) 2.0$ kHz), where the
rf field induced coupling term is~\cite{Suominen98,Vitanov97} 
\begin{equation}
   {\cal H}=\hbar\left( \begin{array}{ccccc}
   0 & \Omega & 0 & 0 & 0 \\
   \Omega&0&\sqrt{\frac{3}{2}} \Omega &0&0\\
   0 & \sqrt{\frac{3}{2}}\Omega & 0 & \sqrt{\frac{3}{2}}\Omega & 0 \\
   0&0&\sqrt{\frac{3}{2}}\Omega&0&\Omega\\
   0&0&0&\Omega&0
   \end{array}\right)
   \quad
   \begin{array}{l}
      |2,-2\rangle \\
      |2,-1\rangle \\
      |2,0\rangle \\
      |2,1\rangle \\
      |2,2\rangle
   \end{array},
   \label{Hamilton}
\end{equation}
in the $|F,M_F\rangle$ basis as indicated.

The wave packet simulations were performed in the same manner as in the
previous study~\cite{Suominen98}. Our initial wave packet has a Gaussian
shape, with a width of $10\ \mu$m. For all practical purposes this wave packet
is very narrow both in position and momentum, and the spreading due to its
natural dispersion is not an important factor. We identify the mean momentum of
the wave packet with the atomic kinetic energy $E_{\rm kin}$, and set 
$E_{\rm kin}/k_B=30\ \mu$K. In the experiment by Desruelle et al.~one had 
typically $B_0=B_{\rm bias}=0.0150$~T, which sets the kinetic energy for 
reaching the resonance points (for any practical value of $\nu_{\rm rf}$) too 
large for realistic numerical simulations. Thus we have introduced
the compensation field and limit $B_0$ to values below $0.0050$~T. But the main
conclusions from our study apply to larger values of $B_0$ and $E_{\rm kin}$,
and many of the results can be scaled to other parameter regions with the
semiclassical models.

Another simplification is that we consider only one spatial dimension. This is
necessary simply because we have chosen to work with relatively large
energies, such as 30 $\mu$K. Numerical wave packet calculations at the
corresponding velocities require on the order of 100 000 points for
both the spatial and temporal dimensions. As discussed in
Ref.~\cite{Suominen98}, however, this is not a crucial simplification.

Basically, we solve the five-component Schr\"odinger equation
\begin{equation}
   i\hbar\frac{\partial\Psi(x,t)}{\partial t} =
   {\cal H}(x)\Psi(x,t),\label{Erwin}
\end{equation}
The components of the state vector $\Psi(x,t)$ stand for the time
dependent probability distributions for each $M_F$ state. The off-diagonal 
part of the Hamiltonian ${\cal H}$ is given Eq.~(\ref{Hamilton}).
The diagonal terms are 
\begin{equation}
    -\frac{\hbar^2}{2m}\frac{\partial^2}{\partial x^2}+U_{\rm
    M_F}(x)-M_Fh\nu_{\rm rf},
\end{equation}
where $m$ is the atomic mass and $U_{\rm M_F}(x)$ are the trap potentials as in
Fig.~\ref{fig1}(a). For states $M_F=-2$ and $M_F=-1$ we use absorbing
boundaries, and reflecting ones for the others. The numerical solution method
is the split operator method, with the kinetic term evaluated by the
Crank-Nicholson approach~\cite{Suominen92a,Garraway95}.

\subsection{Semiclassical models}

For small magnetic fields the rf field induced resonances between adjacent
states occur at the same position, $x=x_C$. In this situation the spin-change
probability for atoms which traverse the resonance is given by the multistate
extension~\cite{Vitanov97,Kazansky96} of the two-state Landau-Zener
model~\cite{LZ}. We have earlier shown that for the evaporation in $^{23}$Na 
$F=2$ state at $E_{\rm kin}/k_B=5\ \mu$K and small $B$ this 
model predicts the wave packet results very well~\cite{Suominen98}.

The solution for the multistate problem can be expressed with the solutions to
the two-state Landau-Zener (LZ) model, so we shall begin by discussing the
two-state case first. Let us consider two potentials, $U_1$ and
$U_2$, which intersect at $x=x_C$ and are coupled by $V$. For strong $B$, when
the crossings are well separated in our alkali $F=2$ system, $V$ is equal to
$\hbar\Omega$ or $\sqrt{3/2}\hbar\Omega$, depending which pair of
adjacent states is involved [see Eq.~(\ref{Hamilton})]. 

In addition to the coupling $V$, the relevant factors are the speed $v_C$ of
the wave packet and the slopes of the trapping potentials $U_{\rm M_F}(x)$ at
the crossing. We define
\begin{equation}
   \alpha=\hbar \left|\frac{d(U_{2}-U_1)}{dx}\right|_{x=x_C}.
\end{equation}
The speed of the wave packet enters the problem as we describe the traversing
of the crossing with a simple classical trajectory, $x=v_C(t-t_0)+x_0$. This
allows us to enter the purely time-dependent description where the population
transfer is given by the two-component Schr\"odinger equation
\begin{equation}
   i\hbar\frac{\partial}{\partial t} \left( \begin{array}{c} \Psi_1(t) \\
   \Psi_2(t)\end{array}\right) = \left( \begin{array}{cc}
   0 &V\\ V&\alpha v_C t \end{array} \right)
   \left( \begin{array}{c} \Psi_1(t) \\ \Psi_2(t) \end{array}\right) .
\end{equation}   
This is the original Landau-Zener theory. In this form it is fully quantum and
we can obtain an analytic expression for state populations $P_1$ and $P_2$
after the crossing. If state 1 was the initial state, then
\begin{equation}
   \begin{array}{ll} P_1 &=1-\exp(-\pi\Lambda) \\ P_2& =\exp(-\pi\Lambda)
   \end{array}, \qquad \Lambda =
   \frac{2V^2}{\hbar\alpha v_C}\label{Lambda}.
\end{equation}
Obviously, the Landau-Zener model is only applicable when the total energy is
higher than the bare-state energy at the resonance point. For more details
about applying LZ theory to wave packet dynamics see
Refs.~\cite{Garraway95,Suominen93,Suominen94}.

And now we return to the original multistate problem. According to the
five-state case of the multistate model (see e.g.~Ref.~\cite{Vitanov97}) the
populations $P_{\rm M_F}$ for the untrapped states after one traversal of the
crossing are
\begin{eqnarray}
   P_2 &=& p^4,\nonumber\\ P_1 &=& 4(1-p)p^3,\nonumber\\ P_0 &=&
   6(1-p)^2p^2,\label{NLZ}\\P_{-1} &=& 4(1-p)^3p,\nonumber\\
   P_{-2} &=& (1-p)^4,\nonumber
\end{eqnarray}
where $p=\exp(-\pi\Lambda)$, and $\Lambda$ is defined by setting
$V=\hbar\Omega/2$. This assumes that we were intially on state $M_F=2$. We
can see that the final population of the initial state, $P_2$, is equal to
$\exp[-\pi\hbar\Omega^2/(2\alpha v_C)]$ for both the two-state and the
multistate model if Hamiltonian~(\ref{Hamilton}) is used. 

\section{Results}\label{results}

\begin{figure}[tbh]
\noindent\centerline{
\psfig{width=90mm,file=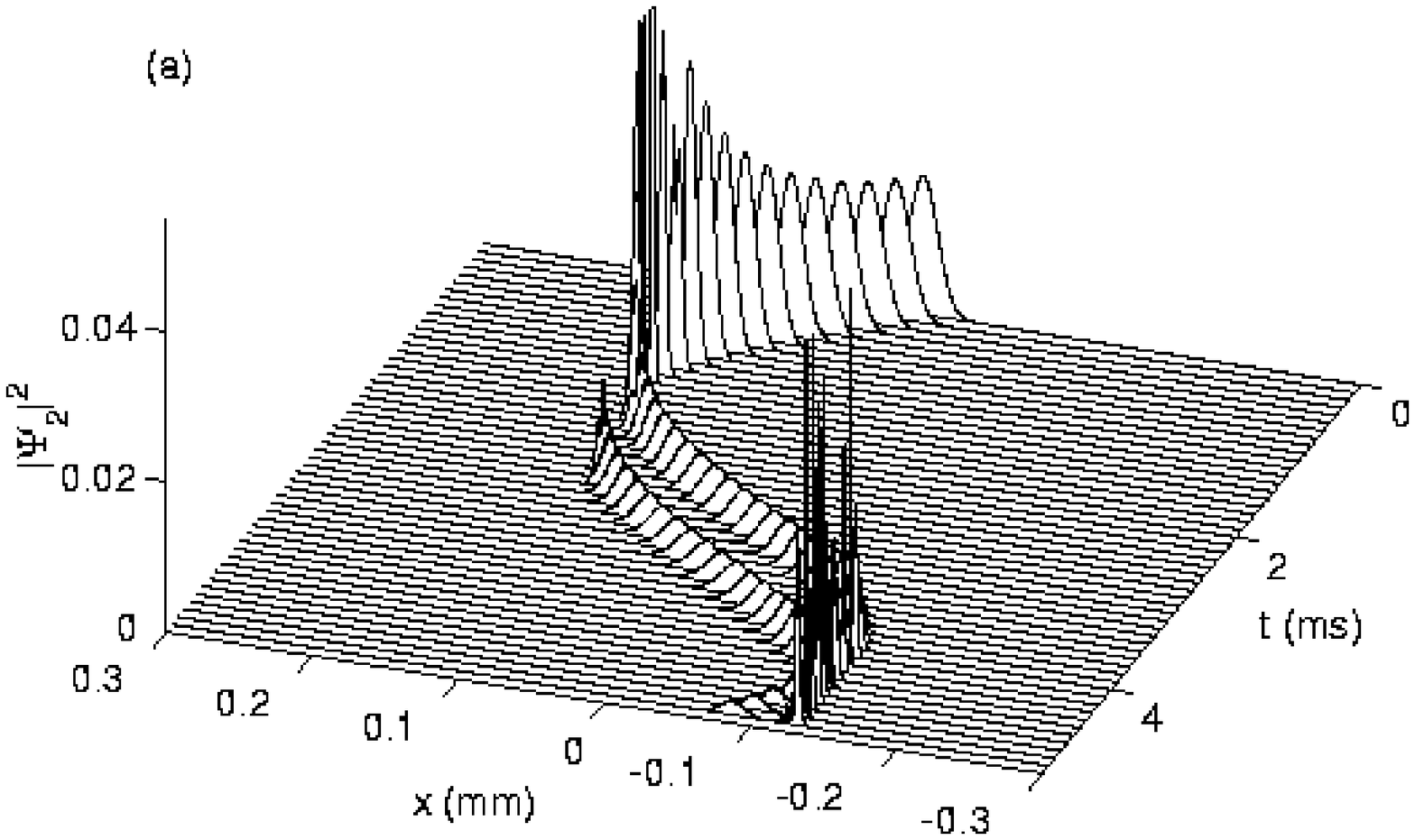}
}

\noindent\centerline{
\psfig{width=90mm,file=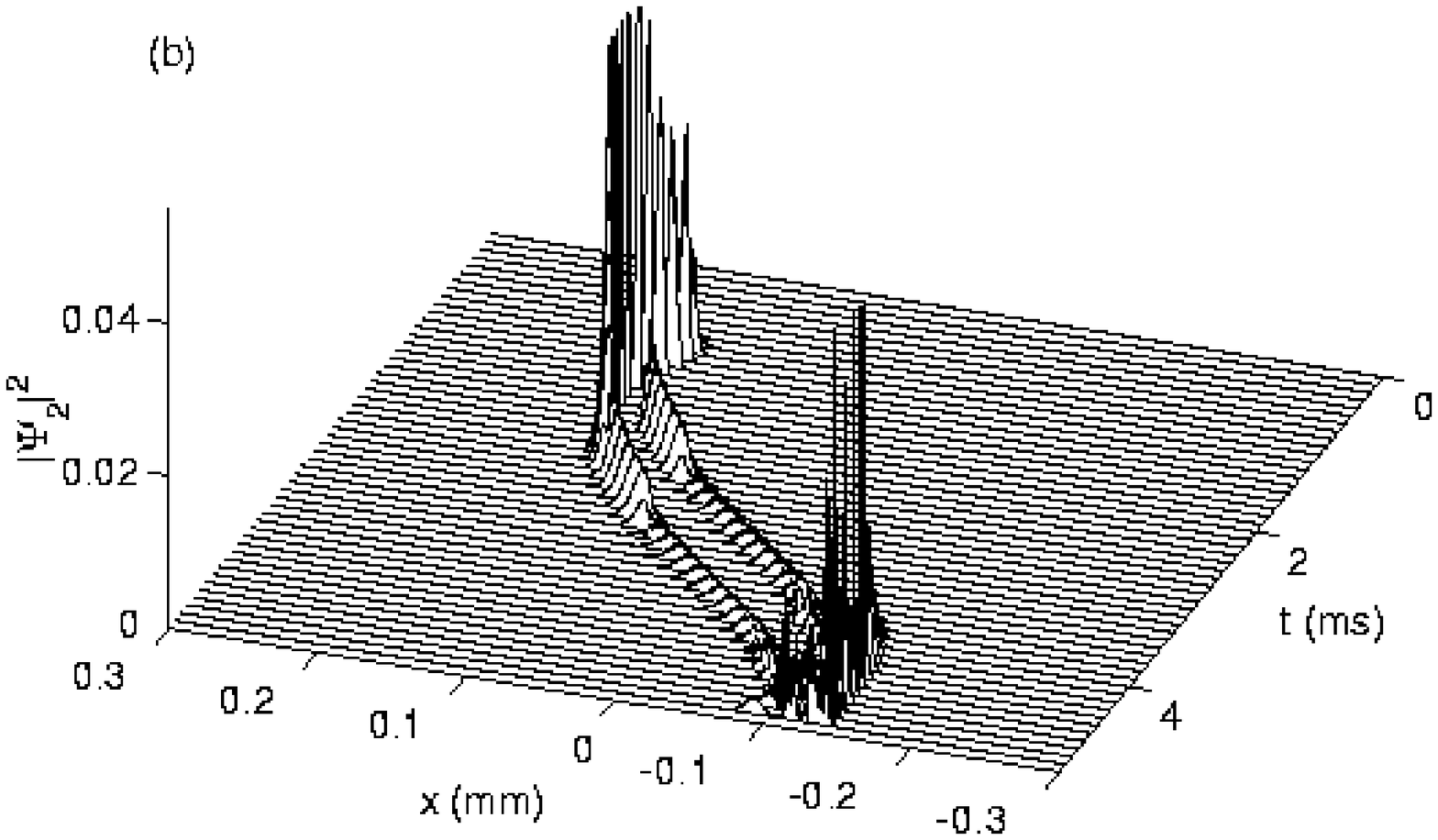}
}

\noindent\centerline{
\psfig{width=90mm,file=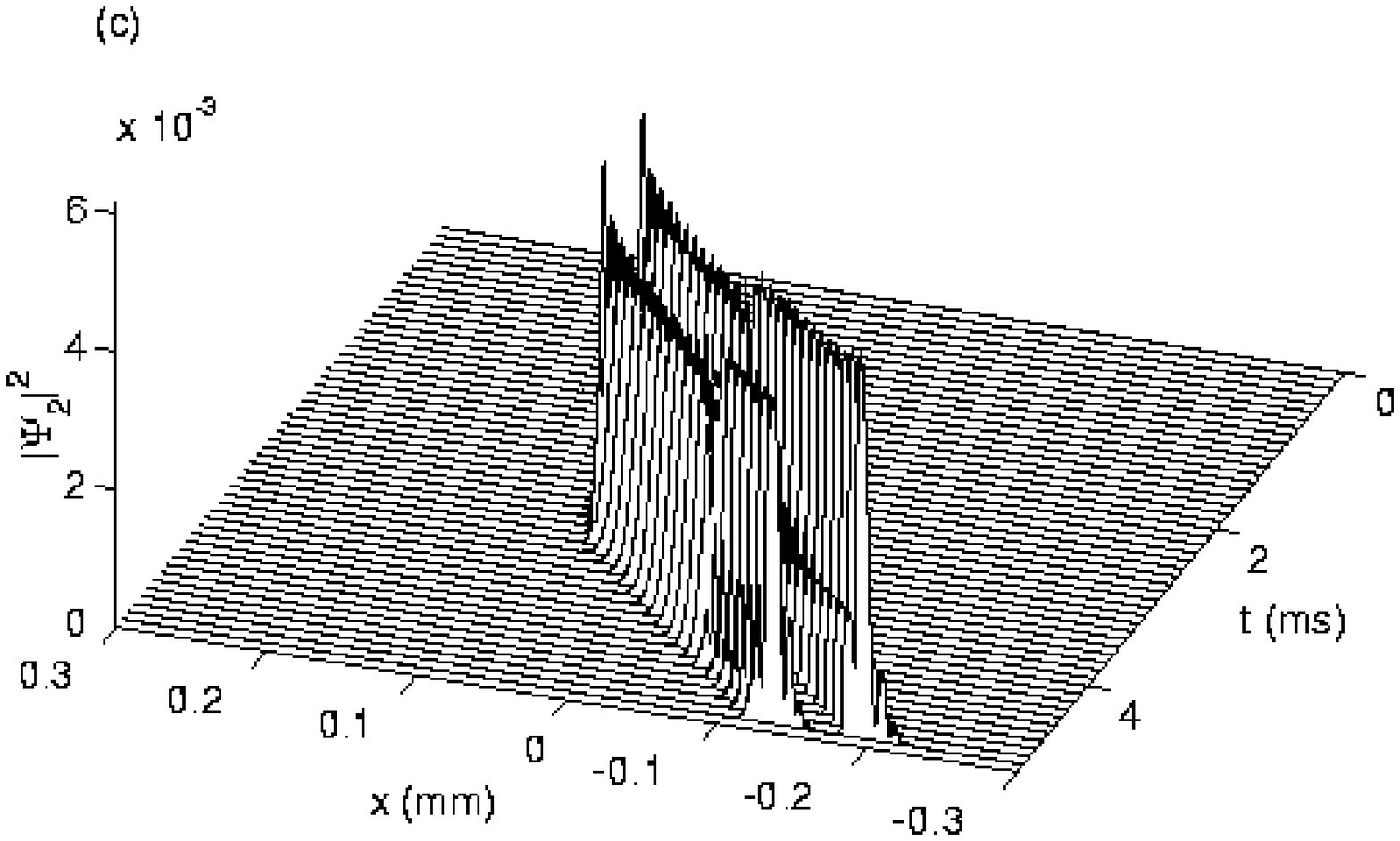}
}
\caption[f7]{The evolution of the wave packet components $|\Psi_{\rm
M_F}(x,t)|^2$ for $^{23}$Na at $B_0=0.0018$ T and $\Omega=(2\pi) 2$ kHz. 
(a) $M_F=2$, (b) $M_F=1$, and (c) $M_F=0$.
\label{fig7}}
\end{figure}

\begin{figure}[htb]
\noindent\centerline{
\psfig{width=90mm,file=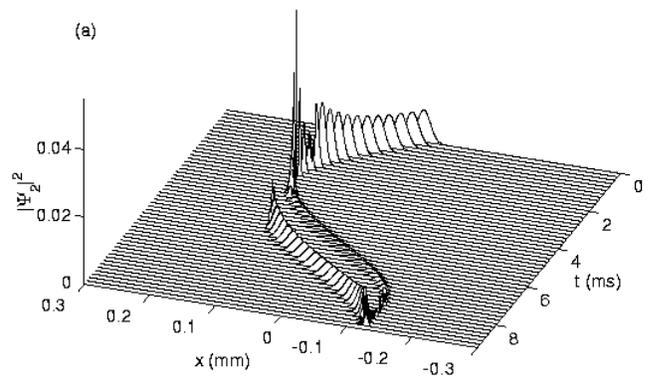}
}
\noindent\centerline{
\psfig{width=90mm,file=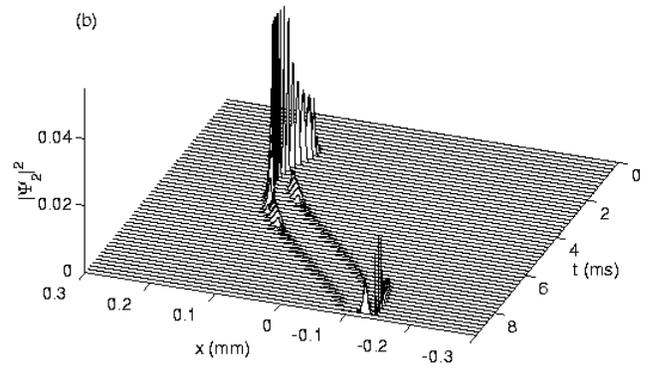}
}
\noindent\centerline{
\psfig{width=90mm,file=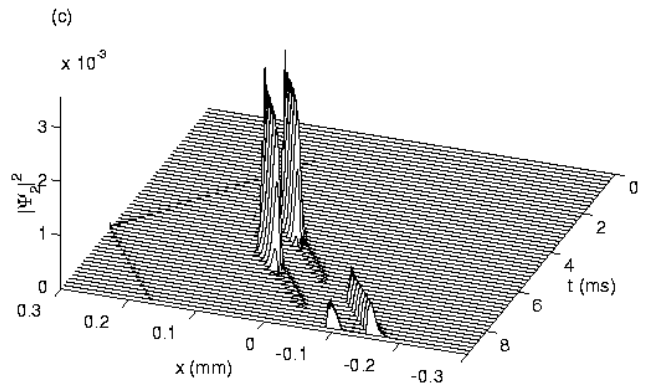}
}
\caption[f8]{The evolution of the wave packet components $|\Psi_{\rm
M_F}(x,t)|^2$ for $^{87}$Rb at $B_0=0.0028$ T and $\Omega=(2\pi) 2$ kHz. 
(a) $M_F=2$, (b) $M_F=1$, and (c) $M_F=0$.
\label{fig8}}
\end{figure}

Typical examples of the atomic wave packet evolution for the three trapping 
states are shown in Figs.~\ref{fig7} and~\ref{fig8}. They demonstrate the 
sloshing discussed e.g.~in Refs.~\cite{Suominen98,Ketterle96,Desruelle99}. 
The amplitudes of the components decrease
as population is partly transferred to another state. Similarly new wave
packet components can appear at crossings. As a wave packet component reaches
a turning point it sharpens strongly. In Fig.~\ref{fig7} we have
$B_0=0.0018$~T, which means that there is no crossing between states $M_F=0$ 
and $M_F=-1$. Population transfer from the state $M_F=1$ to $M_F=0$ is weak.
The $M_F=0$ wave packet component has turning points beyond the integration
space. 

As sloshing continues St\"uckelberg oscillations could take place as split wave
packet components merge again at crossings and interfere (for further 
discussion, see Refs.~\cite{Garraway95,Garraway92}). However, the wave packet 
contains several momentum components and thus such oscillations are not likely 
observed, because they are very sensitive to phase differences. In our
simulations we saw no major indication of inteferences.

\begin{figure}[htb]
\noindent\centerline{
\psfig{width=90mm,file=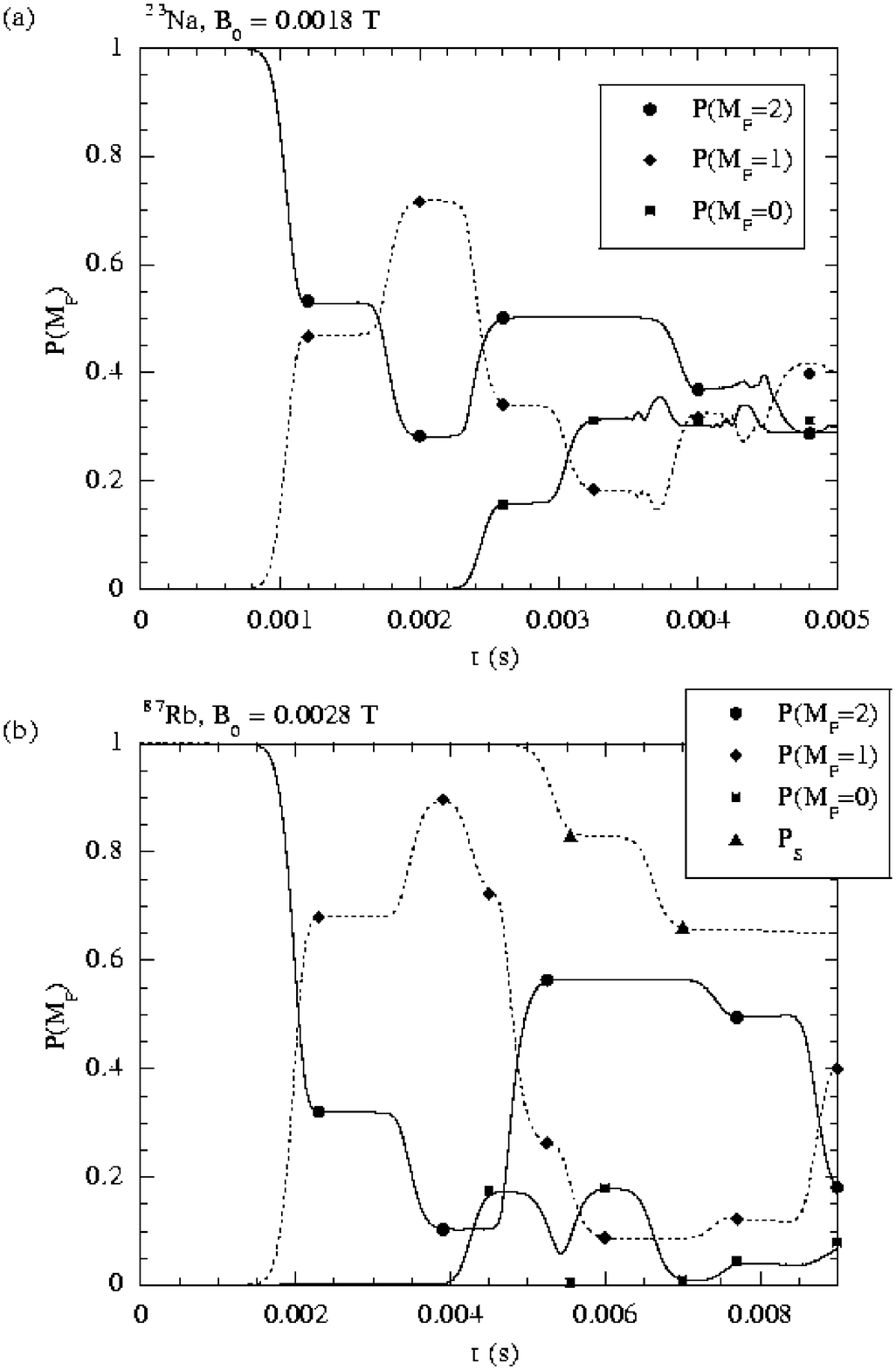}
}
\caption[f9]{The time evolution of the trap state populations.
\label{fig9}}
\end{figure}

In Fig.~\ref{fig9} we track the trap state populations and their sum as the
wave packet sloshes in the trap and traverses several crossings. The magnetic
field values are strong enough to ensure that the crossings are well
separated. We can identify when the various crossings take place although some
of them happen simultaneously. The filled symbols indicate the corresponding
Landau-Zener predictions, and we find that the agreement is excellent. Some
oscillations appear for the $^{23}$Na case [Fig.~\ref{fig7}(a)] at times 
between 3.5
ms and 4.5 ms. These may arise from St\"uckelberg oscillations, but they do not
affect the final transition probabilities, supporting our assumption that in
the end such oscillations average out. Note that for $^{23}$Na there is no 
resonance between states $M_F=0$ and $M_F=-1$, but for $^{87}$Rb there is 
and it is 
seen as a stepwise reduction of $P_S\equiv P(M_F=2)+P(M_F=1)+P(M_F=0)$. 

\begin{figure}[htb]
\noindent\centerline{
\psfig{width=90mm,file=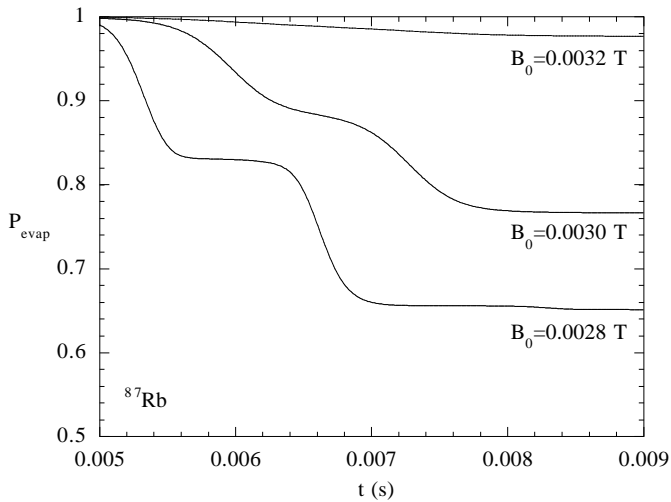}
}
\vspace*{0mm}
\caption[f10]{The time evolution of the trapped population for $^{87}$Rb near
the critical field strength $B_{\rm cr}$.
\label{fig10}}
\end{figure}

Near the critical field $B_{\rm cr}$ the probability to leave the trap via 
states $M_F=-2$ and $M_F=-1$ varies strongly with $B_0$. When $B_0<B_{\rm cr}$
the wave packet meets two crossings between the states $M_F=-1$ and $M_F=0$ as 
it traverses the region around the trap center $x=0$ on state $M_F=0$. 
At both crossings some population leaks into the state $M_F=-1$, as seen in 
Fig.~\ref{fig10} for $B_0=0.0028$~T. As $B_0$ increases, the two crossing 
points, on opposite sides of $x=0$, begin to merge, until they disappear at 
$x=0$ when $B=B_{\rm cr}$. Then the transfer between the two states becomes
off-resonant (or tunnelling), and its probability decreases exponentially as a
function of some ratio of $\Omega$ and the energy difference between the
states $M_F=-1$ and $M_F=0$ at $x=0$. This situation
corresponds to the parabolic level crossing model~\cite{Suominen92}. But the
main point is that the off-resonant process is unlikely to play any major
role.
  
\begin{figure}[htb]
\noindent\centerline{
\psfig{width=90mm,file=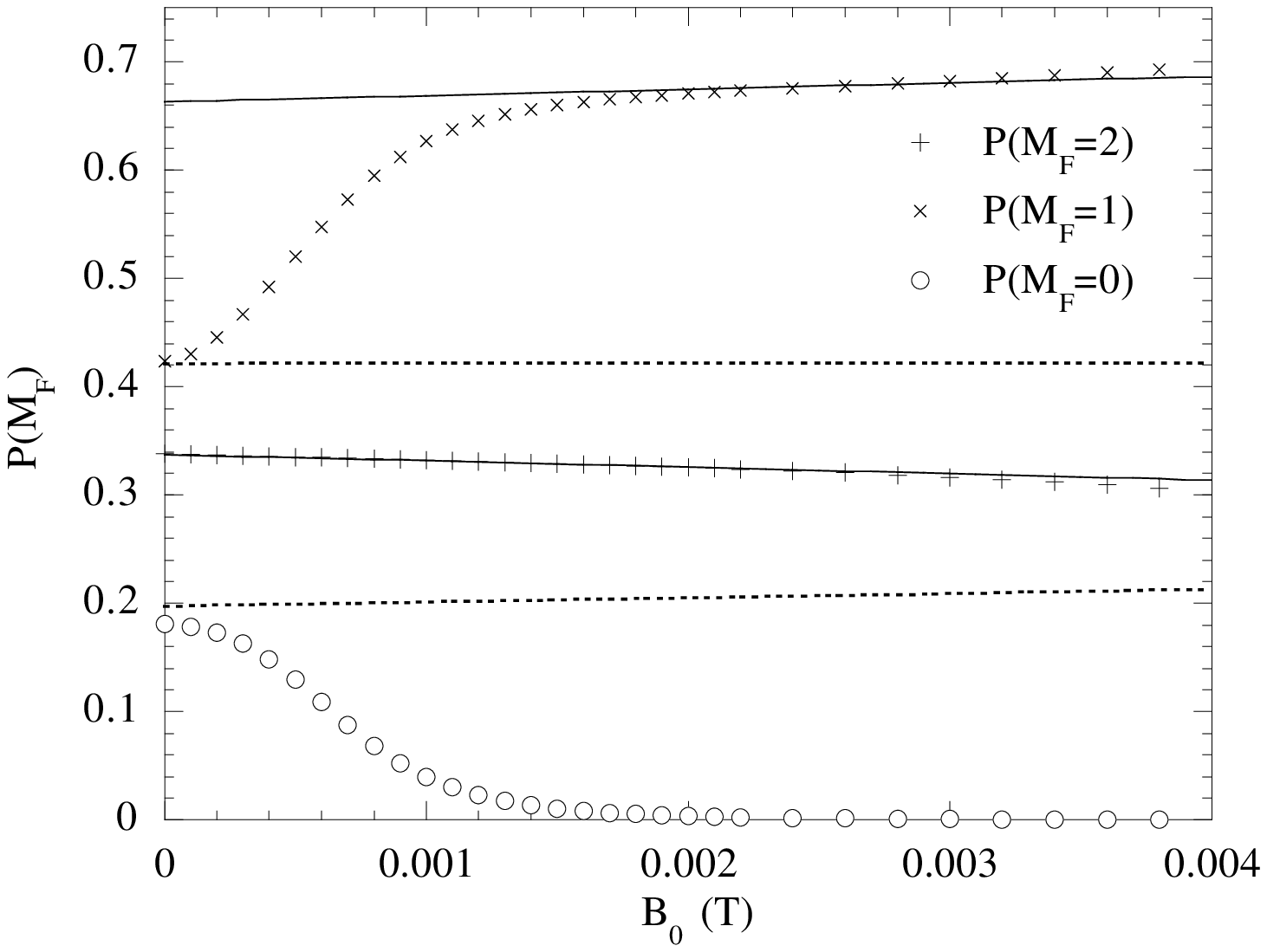}
}
\vspace*{0mm}
\caption[f11]{The population transfer after the first crossing for $^{87}$Rb
as a function of $B_0$. The solid lines are the semiclassical two-state
predictions, and the dotted ones are the multistate model predictions. The
symbols represent the wave packet results.
\label{fig11}}
\end{figure}

Finally, in Fig.~\ref{fig11} we show how the transfer probability between the
trap states at the first crossing changes as a function of $B_0$. The
multistate process transforms smoothly into a two-state process between the
states $M_F=2$ and $M_F=1$. The transition zone is rather large, though,
with $B_0$ ranging from 0 to 0.0010~T. The transfer process in this zone is 
the off-resonant two-photon transfer demonstrated in Fig.~\ref{fig2}. 
An analogous process can occur in atoms interacting with chirped 
pulses~\cite{Maas99}.

An interesting point is that the population of the initial state is not
affected by the fact how the transferred population is distributed to the
other involved states. This seems to be typical for the Landau-Zener 
crossings~\cite{Akulin92}. The solid lines indicate the predictions of the
two-state model, and the dotted lines the multistate model. They change with 
$B_0$ because the location of the first crossing point and thus the wave packet
speed $v_C$ at this point change slightly with $B_0$.

\section{Conclusions}\label{conclusions}

Our results show that in general the semiclassical level crossing models offer
a clear understanding of the single atomic dynamics during the evaporation
process. Also, we have verified with wave packet calculations that the
interpretations presented by Desruelle et al.~for their $^{87}$Rb
experiment~\cite{Desruelle99} are correct. The simple picture of evaporation at
near-zero magnetic fields transforms into a complex sequence of two-state
crossings at field strengths above about 0.0010~T. For all alkali systems
where $F=2$ is the upper hyperfine ground state the evaporation will stop
before condensation as the necessary resonances disappear too soon as a
function of the rf field frequency. We have shown that tunnelling does not
really play a role once the resonances have been lost. Further complications
arise from the fact that the $M_F=0$ state becomes a trapping state. 

In experiments, as suggested by Desruelle et al., one could avoid the problem 
by coupling the $F=2, M_F=2$ trapping state to the $F=1,M_F=1$ nontrapping
state, or by using several rf fields of different frequencies within the
$F=2$ hyperfine manifold.  Although for $^{87}$Rb
one has observed a long-lasting coexistence of $F=1$ and $F=2$ condensates,
theoretical studies~\cite{Julienne97} predict this difficult for $^{23}$Na due
to destructive collisions. Thus the first approach may apply better for 
$^{87}$Rb than for $^{23}$Na.

We have calculated earlier~\cite{Suominen98} that for $^{23}$Na the collisions
between atoms in the $M_F=0$ and $M_F=2$ states are very destructive, with a
rate coefficient on the order of $10^{-11}$ cm$^3$/s. For practical bias field
strengths the $M_F$ state is also a trapping state. Thus the efficiency of
evaporation is reduced, and the time the atoms spend on the $M_F=0$ state
increase, making it more likely to have a destructive, energy releasing
collision. So far condensation on the $F=2$ state for Na has not been achieved.
Even in the weak $B$ field case evaporation can produce atoms on $M_F=0$ state
via nonadiabatic transitions. Thus the role of inelastic collisions is expected
to be enhanced for the field strengths considered here. 

Once condensation is reached, however, the nonlinearity of the Zeeman shifts
can be an asset rather than a nuisance. For instance, one could create a new
type of binary condensates by making a selective transfer of part of the
condensate from the $F=2,M_F=2$ state to the $F=2,M_F=1$ state, either by using
resonant or chirped rf field pulses. Alternatively, two rf pulses of different
frequencies or perhaps a single chirped pulse might allow one to transfer the
condenstate from the $F=2,M_F=2$ state to the $F=2,M_F=0$ state and let it
expand normally, without the need to switch the magnetic fields off. Of course,
this would work only when $B$ is so small that the trapping nature of the $M_F$
state is not too strong. 

\acknowledgements

This research has been supported by the Academy of Finland. We thank A. Aspect
and S. Murdoch for valuable discussions and information.

\end{document}